\documentclass[envcountsect]{llncs}
\usepackage[pdftex]{graphicx}
\graphicspath{{figures/}}
\DeclareGraphicsExtensions{.jpg,.png}

\usepackage[caption=false,font=footnotesize]{subfig}

\usepackage[draft]{minted}
    
\usepackage[font={bf}]{caption}

\usepackage{amsmath}
\usepackage{bbm}
\usepackage{stmaryrd}
\usepackage{array}
\usepackage{color}
\usepackage[hyphens]{url}
\usepackage[pdftitle=Title,pdfauthor=Anonymous]{hyperref}

\def\FV@SetLineWidth{%
\if@FV@ResetMargins\else
\advance\leftmargin\@totalleftmargin
\fi
\advance\leftmargin\FV@XLeftMargin\relax
\advance\rightmargin\FV@XRightMargin\relax
\linewidth\hsize
\advance\linewidth-\leftmargin
\advance\linewidth-\rightmargin
\hfuzz\FancyVerbHFuzz\relax}

\makeatletter
\def\PYGdefault@reset{\let\PYGdefault@it=\relax \let\PYGdefault@bf=\relax%
    \let\PYGdefault@ul=\relax \let\PYGdefault@tc=\relax%
    \let\PYGdefault@bc=\relax \let\PYGdefault@ff=\relax}
\def\PYGdefault@tok#1{\csname PYGdefault@tok@#1\endcsname}
\def\PYGdefault@toks#1+{\ifx\relax#1\empty\else%
    \PYGdefault@tok{#1}\expandafter\PYGdefault@toks\fi}
\def\PYGdefault@do#1{\PYGdefault@bc{\PYGdefault@tc{\PYGdefault@ul{%
    \PYGdefault@it{\PYGdefault@bf{\PYGdefault@ff{#1}}}}}}}
\def\PYGdefault#1#2{\PYGdefault@reset\PYGdefault@toks#1+\relax+\PYGdefault@do{#2}}

\expandafter\def\csname PYGdefault@tok@gd\endcsname{\def\PYGdefault@tc##1{\textcolor[rgb]{0.63,0.00,0.00}{##1}}}
\expandafter\def\csname PYGdefault@tok@gu\endcsname{\let\PYGdefault@bf=\textbf\def\PYGdefault@tc##1{\textcolor[rgb]{0.50,0.00,0.50}{##1}}}
\expandafter\def\csname PYGdefault@tok@gt\endcsname{\def\PYGdefault@tc##1{\textcolor[rgb]{0.00,0.27,0.87}{##1}}}
\expandafter\def\csname PYGdefault@tok@gs\endcsname{\let\PYGdefault@bf=\textbf}
\expandafter\def\csname PYGdefault@tok@gr\endcsname{\def\PYGdefault@tc##1{\textcolor[rgb]{1.00,0.00,0.00}{##1}}}
\expandafter\def\csname PYGdefault@tok@cm\endcsname{\let\PYGdefault@it=\textit\def\PYGdefault@tc##1{\textcolor[rgb]{0.25,0.50,0.50}{##1}}}
\expandafter\def\csname PYGdefault@tok@vg\endcsname{\def\PYGdefault@tc##1{\textcolor[rgb]{0.10,0.09,0.49}{##1}}}
\expandafter\def\csname PYGdefault@tok@vi\endcsname{\def\PYGdefault@tc##1{\textcolor[rgb]{0.10,0.09,0.49}{##1}}}
\expandafter\def\csname PYGdefault@tok@vm\endcsname{\def\PYGdefault@tc##1{\textcolor[rgb]{0.10,0.09,0.49}{##1}}}
\expandafter\def\csname PYGdefault@tok@mh\endcsname{\def\PYGdefault@tc##1{\textcolor[rgb]{0.40,0.40,0.40}{##1}}}
\expandafter\def\csname PYGdefault@tok@cs\endcsname{\let\PYGdefault@it=\textit\def\PYGdefault@tc##1{\textcolor[rgb]{0.25,0.50,0.50}{##1}}}
\expandafter\def\csname PYGdefault@tok@ge\endcsname{\let\PYGdefault@it=\textit}
\expandafter\def\csname PYGdefault@tok@vc\endcsname{\def\PYGdefault@tc##1{\textcolor[rgb]{0.10,0.09,0.49}{##1}}}
\expandafter\def\csname PYGdefault@tok@il\endcsname{\def\PYGdefault@tc##1{\textcolor[rgb]{0.40,0.40,0.40}{##1}}}
\expandafter\def\csname PYGdefault@tok@go\endcsname{\def\PYGdefault@tc##1{\textcolor[rgb]{0.53,0.53,0.53}{##1}}}
\expandafter\def\csname PYGdefault@tok@cp\endcsname{\def\PYGdefault@tc##1{\textcolor[rgb]{0.74,0.48,0.00}{##1}}}
\expandafter\def\csname PYGdefault@tok@gi\endcsname{\def\PYGdefault@tc##1{\textcolor[rgb]{0.00,0.63,0.00}{##1}}}
\expandafter\def\csname PYGdefault@tok@gh\endcsname{\let\PYGdefault@bf=\textbf\def\PYGdefault@tc##1{\textcolor[rgb]{0.00,0.00,0.50}{##1}}}
\expandafter\def\csname PYGdefault@tok@ni\endcsname{\let\PYGdefault@bf=\textbf\def\PYGdefault@tc##1{\textcolor[rgb]{0.60,0.60,0.60}{##1}}}
\expandafter\def\csname PYGdefault@tok@nl\endcsname{\def\PYGdefault@tc##1{\textcolor[rgb]{0.63,0.63,0.00}{##1}}}
\expandafter\def\csname PYGdefault@tok@nn\endcsname{\let\PYGdefault@bf=\textbf\def\PYGdefault@tc##1{\textcolor[rgb]{0.00,0.00,1.00}{##1}}}
\expandafter\def\csname PYGdefault@tok@no\endcsname{\def\PYGdefault@tc##1{\textcolor[rgb]{0.53,0.00,0.00}{##1}}}
\expandafter\def\csname PYGdefault@tok@na\endcsname{\def\PYGdefault@tc##1{\textcolor[rgb]{0.49,0.56,0.16}{##1}}}
\expandafter\def\csname PYGdefault@tok@nb\endcsname{\def\PYGdefault@tc##1{\textcolor[rgb]{0.00,0.50,0.00}{##1}}}
\expandafter\def\csname PYGdefault@tok@nc\endcsname{\let\PYGdefault@bf=\textbf\def\PYGdefault@tc##1{\textcolor[rgb]{0.00,0.00,1.00}{##1}}}
\expandafter\def\csname PYGdefault@tok@nd\endcsname{\def\PYGdefault@tc##1{\textcolor[rgb]{0.67,0.13,1.00}{##1}}}
\expandafter\def\csname PYGdefault@tok@ne\endcsname{\let\PYGdefault@bf=\textbf\def\PYGdefault@tc##1{\textcolor[rgb]{0.82,0.25,0.23}{##1}}}
\expandafter\def\csname PYGdefault@tok@nf\endcsname{\def\PYGdefault@tc##1{\textcolor[rgb]{0.00,0.00,1.00}{##1}}}
\expandafter\def\csname PYGdefault@tok@si\endcsname{\let\PYGdefault@bf=\textbf\def\PYGdefault@tc##1{\textcolor[rgb]{0.73,0.40,0.53}{##1}}}
\expandafter\def\csname PYGdefault@tok@s2\endcsname{\def\PYGdefault@tc##1{\textcolor[rgb]{0.73,0.13,0.13}{##1}}}
\expandafter\def\csname PYGdefault@tok@nt\endcsname{\let\PYGdefault@bf=\textbf\def\PYGdefault@tc##1{\textcolor[rgb]{0.00,0.50,0.00}{##1}}}
\expandafter\def\csname PYGdefault@tok@nv\endcsname{\def\PYGdefault@tc##1{\textcolor[rgb]{0.10,0.09,0.49}{##1}}}
\expandafter\def\csname PYGdefault@tok@s1\endcsname{\def\PYGdefault@tc##1{\textcolor[rgb]{0.73,0.13,0.13}{##1}}}
\expandafter\def\csname PYGdefault@tok@dl\endcsname{\def\PYGdefault@tc##1{\textcolor[rgb]{0.73,0.13,0.13}{##1}}}
\expandafter\def\csname PYGdefault@tok@ch\endcsname{\let\PYGdefault@it=\textit\def\PYGdefault@tc##1{\textcolor[rgb]{0.25,0.50,0.50}{##1}}}
\expandafter\def\csname PYGdefault@tok@m\endcsname{\def\PYGdefault@tc##1{\textcolor[rgb]{0.40,0.40,0.40}{##1}}}
\expandafter\def\csname PYGdefault@tok@gp\endcsname{\let\PYGdefault@bf=\textbf\def\PYGdefault@tc##1{\textcolor[rgb]{0.00,0.00,0.50}{##1}}}
\expandafter\def\csname PYGdefault@tok@sh\endcsname{\def\PYGdefault@tc##1{\textcolor[rgb]{0.73,0.13,0.13}{##1}}}
\expandafter\def\csname PYGdefault@tok@ow\endcsname{\let\PYGdefault@bf=\textbf\def\PYGdefault@tc##1{\textcolor[rgb]{0.67,0.13,1.00}{##1}}}
\expandafter\def\csname PYGdefault@tok@sx\endcsname{\def\PYGdefault@tc##1{\textcolor[rgb]{0.00,0.50,0.00}{##1}}}
\expandafter\def\csname PYGdefault@tok@bp\endcsname{\def\PYGdefault@tc##1{\textcolor[rgb]{0.00,0.50,0.00}{##1}}}
\expandafter\def\csname PYGdefault@tok@c1\endcsname{\let\PYGdefault@it=\textit\def\PYGdefault@tc##1{\textcolor[rgb]{0.25,0.50,0.50}{##1}}}
\expandafter\def\csname PYGdefault@tok@fm\endcsname{\def\PYGdefault@tc##1{\textcolor[rgb]{0.00,0.00,1.00}{##1}}}
\expandafter\def\csname PYGdefault@tok@o\endcsname{\def\PYGdefault@tc##1{\textcolor[rgb]{0.40,0.40,0.40}{##1}}}
\expandafter\def\csname PYGdefault@tok@kc\endcsname{\let\PYGdefault@bf=\textbf\def\PYGdefault@tc##1{\textcolor[rgb]{0.00,0.50,0.00}{##1}}}
\expandafter\def\csname PYGdefault@tok@c\endcsname{\let\PYGdefault@it=\textit\def\PYGdefault@tc##1{\textcolor[rgb]{0.25,0.50,0.50}{##1}}}
\expandafter\def\csname PYGdefault@tok@mf\endcsname{\def\PYGdefault@tc##1{\textcolor[rgb]{0.40,0.40,0.40}{##1}}}
\expandafter\def\csname PYGdefault@tok@err\endcsname{\def\PYGdefault@bc##1{\setlength{\fboxsep}{0pt}\fcolorbox[rgb]{1.00,0.00,0.00}{1,1,1}{\strut ##1}}}
\expandafter\def\csname PYGdefault@tok@mb\endcsname{\def\PYGdefault@tc##1{\textcolor[rgb]{0.40,0.40,0.40}{##1}}}
\expandafter\def\csname PYGdefault@tok@ss\endcsname{\def\PYGdefault@tc##1{\textcolor[rgb]{0.10,0.09,0.49}{##1}}}
\expandafter\def\csname PYGdefault@tok@sr\endcsname{\def\PYGdefault@tc##1{\textcolor[rgb]{0.73,0.40,0.53}{##1}}}
\expandafter\def\csname PYGdefault@tok@mo\endcsname{\def\PYGdefault@tc##1{\textcolor[rgb]{0.40,0.40,0.40}{##1}}}
\expandafter\def\csname PYGdefault@tok@kd\endcsname{\let\PYGdefault@bf=\textbf\def\PYGdefault@tc##1{\textcolor[rgb]{0.00,0.50,0.00}{##1}}}
\expandafter\def\csname PYGdefault@tok@mi\endcsname{\def\PYGdefault@tc##1{\textcolor[rgb]{0.40,0.40,0.40}{##1}}}
\expandafter\def\csname PYGdefault@tok@kn\endcsname{\let\PYGdefault@bf=\textbf\def\PYGdefault@tc##1{\textcolor[rgb]{0.00,0.50,0.00}{##1}}}
\expandafter\def\csname PYGdefault@tok@cpf\endcsname{\let\PYGdefault@it=\textit\def\PYGdefault@tc##1{\textcolor[rgb]{0.25,0.50,0.50}{##1}}}
\expandafter\def\csname PYGdefault@tok@kr\endcsname{\let\PYGdefault@bf=\textbf\def\PYGdefault@tc##1{\textcolor[rgb]{0.00,0.50,0.00}{##1}}}
\expandafter\def\csname PYGdefault@tok@s\endcsname{\def\PYGdefault@tc##1{\textcolor[rgb]{0.73,0.13,0.13}{##1}}}
\expandafter\def\csname PYGdefault@tok@kp\endcsname{\def\PYGdefault@tc##1{\textcolor[rgb]{0.00,0.50,0.00}{##1}}}
\expandafter\def\csname PYGdefault@tok@w\endcsname{\def\PYGdefault@tc##1{\textcolor[rgb]{0.73,0.73,0.73}{##1}}}
\expandafter\def\csname PYGdefault@tok@kt\endcsname{\def\PYGdefault@tc##1{\textcolor[rgb]{0.69,0.00,0.25}{##1}}}
\expandafter\def\csname PYGdefault@tok@sc\endcsname{\def\PYGdefault@tc##1{\textcolor[rgb]{0.73,0.13,0.13}{##1}}}
\expandafter\def\csname PYGdefault@tok@sb\endcsname{\def\PYGdefault@tc##1{\textcolor[rgb]{0.73,0.13,0.13}{##1}}}
\expandafter\def\csname PYGdefault@tok@sa\endcsname{\def\PYGdefault@tc##1{\textcolor[rgb]{0.73,0.13,0.13}{##1}}}
\expandafter\def\csname PYGdefault@tok@k\endcsname{\let\PYGdefault@bf=\textbf\def\PYGdefault@tc##1{\textcolor[rgb]{0.00,0.50,0.00}{##1}}}
\expandafter\def\csname PYGdefault@tok@se\endcsname{\let\PYGdefault@bf=\textbf\def\PYGdefault@tc##1{\textcolor[rgb]{0.73,0.40,0.13}{##1}}}
\expandafter\def\csname PYGdefault@tok@sd\endcsname{\let\PYGdefault@it=\textit\def\PYGdefault@tc##1{\textcolor[rgb]{0.73,0.13,0.13}{##1}}}

\makeatother

\makeatletter
\def\PYG@reset{\let\PYG@it=\relax \let\PYG@bf=\relax%
    \let\PYG@ul=\relax \let\PYG@tc=\relax%
    \let\PYG@bc=\relax \let\PYG@ff=\relax}
\def\PYG@tok#1{\csname PYG@tok@#1\endcsname}
\def\PYG@toks#1+{\ifx\relax#1\empty\else%
    \PYG@tok{#1}\expandafter\PYG@toks\fi}
\def\PYG@do#1{\PYG@bc{\PYG@tc{\PYG@ul{%
    \PYG@it{\PYG@bf{\PYG@ff{#1}}}}}}}
\def\PYG#1#2{\PYG@reset\PYG@toks#1+\relax+\PYG@do{#2}}

\expandafter\def\csname PYG@tok@gd\endcsname{\def\PYG@tc##1{\textcolor[rgb]{0.63,0.00,0.00}{##1}}}
\expandafter\def\csname PYG@tok@gu\endcsname{\let\PYG@bf=\textbf\def\PYG@tc##1{\textcolor[rgb]{0.50,0.00,0.50}{##1}}}
\expandafter\def\csname PYG@tok@gt\endcsname{\def\PYG@tc##1{\textcolor[rgb]{0.00,0.27,0.87}{##1}}}
\expandafter\def\csname PYG@tok@gs\endcsname{\let\PYG@bf=\textbf}
\expandafter\def\csname PYG@tok@gr\endcsname{\def\PYG@tc##1{\textcolor[rgb]{1.00,0.00,0.00}{##1}}}
\expandafter\def\csname PYG@tok@cm\endcsname{\let\PYG@it=\textit\def\PYG@tc##1{\textcolor[rgb]{0.25,0.50,0.50}{##1}}}
\expandafter\def\csname PYG@tok@vg\endcsname{\def\PYG@tc##1{\textcolor[rgb]{0.10,0.09,0.49}{##1}}}
\expandafter\def\csname PYG@tok@vi\endcsname{\def\PYG@tc##1{\textcolor[rgb]{0.10,0.09,0.49}{##1}}}
\expandafter\def\csname PYG@tok@vm\endcsname{\def\PYG@tc##1{\textcolor[rgb]{0.10,0.09,0.49}{##1}}}
\expandafter\def\csname PYG@tok@mh\endcsname{\def\PYG@tc##1{\textcolor[rgb]{0.40,0.40,0.40}{##1}}}
\expandafter\def\csname PYG@tok@cs\endcsname{\let\PYG@it=\textit\def\PYG@tc##1{\textcolor[rgb]{0.25,0.50,0.50}{##1}}}
\expandafter\def\csname PYG@tok@ge\endcsname{\let\PYG@it=\textit}
\expandafter\def\csname PYG@tok@vc\endcsname{\def\PYG@tc##1{\textcolor[rgb]{0.10,0.09,0.49}{##1}}}
\expandafter\def\csname PYG@tok@il\endcsname{\def\PYG@tc##1{\textcolor[rgb]{0.40,0.40,0.40}{##1}}}
\expandafter\def\csname PYG@tok@go\endcsname{\def\PYG@tc##1{\textcolor[rgb]{0.53,0.53,0.53}{##1}}}
\expandafter\def\csname PYG@tok@cp\endcsname{\def\PYG@tc##1{\textcolor[rgb]{0.74,0.48,0.00}{##1}}}
\expandafter\def\csname PYG@tok@gi\endcsname{\def\PYG@tc##1{\textcolor[rgb]{0.00,0.63,0.00}{##1}}}
\expandafter\def\csname PYG@tok@gh\endcsname{\let\PYG@bf=\textbf\def\PYG@tc##1{\textcolor[rgb]{0.00,0.00,0.50}{##1}}}
\expandafter\def\csname PYG@tok@ni\endcsname{\let\PYG@bf=\textbf\def\PYG@tc##1{\textcolor[rgb]{0.60,0.60,0.60}{##1}}}
\expandafter\def\csname PYG@tok@nl\endcsname{\def\PYG@tc##1{\textcolor[rgb]{0.63,0.63,0.00}{##1}}}
\expandafter\def\csname PYG@tok@nn\endcsname{\let\PYG@bf=\textbf\def\PYG@tc##1{\textcolor[rgb]{0.00,0.00,1.00}{##1}}}
\expandafter\def\csname PYG@tok@no\endcsname{\def\PYG@tc##1{\textcolor[rgb]{0.53,0.00,0.00}{##1}}}
\expandafter\def\csname PYG@tok@na\endcsname{\def\PYG@tc##1{\textcolor[rgb]{0.49,0.56,0.16}{##1}}}
\expandafter\def\csname PYG@tok@nb\endcsname{\def\PYG@tc##1{\textcolor[rgb]{0.00,0.50,0.00}{##1}}}
\expandafter\def\csname PYG@tok@nc\endcsname{\let\PYG@bf=\textbf\def\PYG@tc##1{\textcolor[rgb]{0.00,0.00,1.00}{##1}}}
\expandafter\def\csname PYG@tok@nd\endcsname{\def\PYG@tc##1{\textcolor[rgb]{0.67,0.13,1.00}{##1}}}
\expandafter\def\csname PYG@tok@ne\endcsname{\let\PYG@bf=\textbf\def\PYG@tc##1{\textcolor[rgb]{0.82,0.25,0.23}{##1}}}
\expandafter\def\csname PYG@tok@nf\endcsname{\def\PYG@tc##1{\textcolor[rgb]{0.00,0.00,1.00}{##1}}}
\expandafter\def\csname PYG@tok@si\endcsname{\let\PYG@bf=\textbf\def\PYG@tc##1{\textcolor[rgb]{0.73,0.40,0.53}{##1}}}
\expandafter\def\csname PYG@tok@s2\endcsname{\def\PYG@tc##1{\textcolor[rgb]{0.73,0.13,0.13}{##1}}}
\expandafter\def\csname PYG@tok@nt\endcsname{\let\PYG@bf=\textbf\def\PYG@tc##1{\textcolor[rgb]{0.00,0.50,0.00}{##1}}}
\expandafter\def\csname PYG@tok@nv\endcsname{\def\PYG@tc##1{\textcolor[rgb]{0.10,0.09,0.49}{##1}}}
\expandafter\def\csname PYG@tok@s1\endcsname{\def\PYG@tc##1{\textcolor[rgb]{0.73,0.13,0.13}{##1}}}
\expandafter\def\csname PYG@tok@dl\endcsname{\def\PYG@tc##1{\textcolor[rgb]{0.73,0.13,0.13}{##1}}}
\expandafter\def\csname PYG@tok@ch\endcsname{\let\PYG@it=\textit\def\PYG@tc##1{\textcolor[rgb]{0.25,0.50,0.50}{##1}}}
\expandafter\def\csname PYG@tok@m\endcsname{\def\PYG@tc##1{\textcolor[rgb]{0.40,0.40,0.40}{##1}}}
\expandafter\def\csname PYG@tok@gp\endcsname{\let\PYG@bf=\textbf\def\PYG@tc##1{\textcolor[rgb]{0.00,0.00,0.50}{##1}}}
\expandafter\def\csname PYG@tok@sh\endcsname{\def\PYG@tc##1{\textcolor[rgb]{0.73,0.13,0.13}{##1}}}
\expandafter\def\csname PYG@tok@ow\endcsname{\let\PYG@bf=\textbf\def\PYG@tc##1{\textcolor[rgb]{0.67,0.13,1.00}{##1}}}
\expandafter\def\csname PYG@tok@sx\endcsname{\def\PYG@tc##1{\textcolor[rgb]{0.00,0.50,0.00}{##1}}}
\expandafter\def\csname PYG@tok@bp\endcsname{\def\PYG@tc##1{\textcolor[rgb]{0.00,0.50,0.00}{##1}}}
\expandafter\def\csname PYG@tok@c1\endcsname{\let\PYG@it=\textit\def\PYG@tc##1{\textcolor[rgb]{0.25,0.50,0.50}{##1}}}
\expandafter\def\csname PYG@tok@fm\endcsname{\def\PYG@tc##1{\textcolor[rgb]{0.00,0.00,1.00}{##1}}}
\expandafter\def\csname PYG@tok@o\endcsname{\def\PYG@tc##1{\textcolor[rgb]{0.40,0.40,0.40}{##1}}}
\expandafter\def\csname PYG@tok@kc\endcsname{\let\PYG@bf=\textbf\def\PYG@tc##1{\textcolor[rgb]{0.00,0.50,0.00}{##1}}}
\expandafter\def\csname PYG@tok@c\endcsname{\let\PYG@it=\textit\def\PYG@tc##1{\textcolor[rgb]{0.25,0.50,0.50}{##1}}}
\expandafter\def\csname PYG@tok@mf\endcsname{\def\PYG@tc##1{\textcolor[rgb]{0.40,0.40,0.40}{##1}}}
\expandafter\def\csname PYG@tok@err\endcsname{\def\PYG@bc##1{\setlength{\fboxsep}{0pt}\fcolorbox[rgb]{1.00,0.00,0.00}{1,1,1}{\strut ##1}}}
\expandafter\def\csname PYG@tok@mb\endcsname{\def\PYG@tc##1{\textcolor[rgb]{0.40,0.40,0.40}{##1}}}
\expandafter\def\csname PYG@tok@ss\endcsname{\def\PYG@tc##1{\textcolor[rgb]{0.10,0.09,0.49}{##1}}}
\expandafter\def\csname PYG@tok@sr\endcsname{\def\PYG@tc##1{\textcolor[rgb]{0.73,0.40,0.53}{##1}}}
\expandafter\def\csname PYG@tok@mo\endcsname{\def\PYG@tc##1{\textcolor[rgb]{0.40,0.40,0.40}{##1}}}
\expandafter\def\csname PYG@tok@kd\endcsname{\let\PYG@bf=\textbf\def\PYG@tc##1{\textcolor[rgb]{0.00,0.50,0.00}{##1}}}
\expandafter\def\csname PYG@tok@mi\endcsname{\def\PYG@tc##1{\textcolor[rgb]{0.40,0.40,0.40}{##1}}}
\expandafter\def\csname PYG@tok@kn\endcsname{\let\PYG@bf=\textbf\def\PYG@tc##1{\textcolor[rgb]{0.00,0.50,0.00}{##1}}}
\expandafter\def\csname PYG@tok@cpf\endcsname{\let\PYG@it=\textit\def\PYG@tc##1{\textcolor[rgb]{0.25,0.50,0.50}{##1}}}
\expandafter\def\csname PYG@tok@kr\endcsname{\let\PYG@bf=\textbf\def\PYG@tc##1{\textcolor[rgb]{0.00,0.50,0.00}{##1}}}
\expandafter\def\csname PYG@tok@s\endcsname{\def\PYG@tc##1{\textcolor[rgb]{0.73,0.13,0.13}{##1}}}
\expandafter\def\csname PYG@tok@kp\endcsname{\def\PYG@tc##1{\textcolor[rgb]{0.00,0.50,0.00}{##1}}}
\expandafter\def\csname PYG@tok@w\endcsname{\def\PYG@tc##1{\textcolor[rgb]{0.73,0.73,0.73}{##1}}}
\expandafter\def\csname PYG@tok@kt\endcsname{\def\PYG@tc##1{\textcolor[rgb]{0.69,0.00,0.25}{##1}}}
\expandafter\def\csname PYG@tok@sc\endcsname{\def\PYG@tc##1{\textcolor[rgb]{0.73,0.13,0.13}{##1}}}
\expandafter\def\csname PYG@tok@sb\endcsname{\def\PYG@tc##1{\textcolor[rgb]{0.73,0.13,0.13}{##1}}}
\expandafter\def\csname PYG@tok@sa\endcsname{\def\PYG@tc##1{\textcolor[rgb]{0.73,0.13,0.13}{##1}}}
\expandafter\def\csname PYG@tok@k\endcsname{\let\PYG@bf=\textbf\def\PYG@tc##1{\textcolor[rgb]{0.00,0.50,0.00}{##1}}}
\expandafter\def\csname PYG@tok@se\endcsname{\let\PYG@bf=\textbf\def\PYG@tc##1{\textcolor[rgb]{0.73,0.40,0.13}{##1}}}
\expandafter\def\csname PYG@tok@sd\endcsname{\let\PYG@it=\textit\def\PYG@tc##1{\textcolor[rgb]{0.73,0.13,0.13}{##1}}}

\def\PYGZus{\char`\_}
\def\PYGZob{\char`\{}
\def\PYGZcb{\char`\}}

\def\PYGZam{\char`\&}
\def\PYGZlt{\char`\<}
\def\PYGZgt{\char`\>}

\def\PYGZhy{\char`\-}

\def\PYGZdq{\char`\"}

\makeatother

\usepackage{xspace}

\newcommand{\eg}{\textit{e.g.,}\xspace}

\newenvironment{compactlistn}
  {\begin{enumerate} 
  \setlength{\itemsep}{0pt} 
  \setlength{\parskip}{0pt}} 
  {\end{enumerate}}

\newcommand{\Velocity}{\textsf{Velocity}\xspace}
\newcommand{\PriceGeth}{\textsf{PriceGeth}\xspace}
\newcommand{\Oraclizeit}{\textsf{Oraclizeit}\xspace}

\newcommand{\PriceGethGit}{\url{https://github.com/VelocityMarket/pricegeth}}

\definecolor{mintedbackground}{rgb}{0.95,0.95,0.95}

\begin{document}
\frontmatter
\mainmatter

\title{\Large \bf On the feasibility of decentralized derivatives markets }
\author{}

\author{
	Shayan Eskandari \inst{1}
	Jeremy Clark \inst{2}
	Vignesh Sundaresan \inst{1}
	Moe Adham \inst{1} 
	}

\institute{
	1 Bitaccess \\*
	2 Concordia University \\*
	}

\maketitle


\begin{abstract}


In this paper, we present Velocity, a decentralized market deployed on Ethereum for trading a custom type of derivative option. To enable the smart contract to work, we also implement a price fetching tool called PriceGeth. We present this as a case study, noting challenges in development of the system that might be of independent interest to whose working on smart contract implementations. We also apply recent academic results on the security of the Solidity smart contract language in validating our code's security. Finally, we discuss more generally the use of smart contracts in modelling financial derivatives.

\end{abstract}


\section{Introductory Remarks}

The introduction of Bitcoin~\cite{nakamoto2008bitcoin} in 2009 led to a new frontier in decentralizing technologies, both in finance and elsewhere. Of the many implementations, we note a few:  file systems like The InterPlanetary File System (IPFS)~\cite{benet2014ipfs}, dynamic name servers like DNSChain~\cite{dnschain} and MaidSafe, a fully distributed platform~\cite{irvine2010maidsafe}. For our purposes, the most interesting technology is Ethereum~\cite{buterin2014next}\cite{wood2014ethereum} --- a decentralized general transaction ledger. Ethereum in simple words is a decentralized computer that can run code, called smart contracts, which enforce the performance of an agreed upon set of negotiated standards in an automated and immutable way. Smart contracts can be designed to disintermediate traditional trusted parties, replacing them with pre-­defined logical parameters. The smart contract concept is not new and was introduced by Szabo in 1997~\cite{szabo1997idea}, however there has not been any real implementation of it until Bitcoin, and then in a much more flexible and verbose fashion: Ethereum.

Under the umbrella of ``fintech'', ``blockchain'', and ``distributed ledger technology'', many legacy entities in the financial world (investment banks, security exchanges, clearinghouses, etc.) have expressed interest (through whitepapers and commercial partnerships and consortiums) in decentralizing financial markets. Derivative markets are often cited as a potential target. From the other end, papers on Ethereum and tutorials on Solidity (a high level programming language for Ethereum) often use derivatives as an example application. So there is a degree of consensus that derivatives running on Ethereum is an interesting application to study, but we are not aware of any public projects to attempt to build a derivative market in a serious way. This paper is a first step in that direction. 
 

\subsection{Scope \& Contributions.} 

A simplification of a derivative is as follows: two parties enter an agreement where the first stands to profit if a specified security (\eg stock) appreciates in value over a specified time-period and the second stands to profit if it falls. Since the profitability of the agreement is derived directly from the price of the security, it is called a derivative instrument. The exact operational details that realize this property differs between types of derivatives. The most common derivative is a put/call option which gives the second party (called the \textit{buyer}) the opportunity (but not obligation) to buy/sell a security at a specified price (\textit{strike price}) at (\textit{American}) or within (\textit{European}) a specified time (\textit{expiration}). The buyer pays the first party (the \textit{seller}) a flat fee (\textit{option price}) when purchasing the option. Derivatives are generally held to hedge risks in price movements or for speculation. 

In a decentralized derivative system, a buyer and seller can have fast and automatic clearing and settlement (straight through processing) of the derivative without trusting a third party. However the design of a market must consider the following challenges:

\begin{compactlistn}
\item \textbf{Terms of the Contract.} The terms of derivative must be expressible in the smart contract language. In this paper, we write contracts in Solidity for the Ethereum blockchain which is sufficient for describing the core aspects of the contract. We present a full implementation stack (from the smart contracts to a UI) for buying/selling a special type of derivative instrument. We pay special attention to common security risks in developing Solidity-based contracts. 
\item \textbf{Counterparty Risk.} In most derivatives, the seller is obliged to buy/sell securities upon request of the buyer subject to the terms of the derivative. A seller might choose to not follow through with her obligations. In a centralized setting, identity, reputation and legal recourse are used to combat this. In a decentralized environment, this problem must be addressed. In this paper (and the reason we position it as a first step), we start with derivatives that are fully collateralized --- meaning the full settlement amount under all outcomes is capped and this amount is locked to the contract at initiation time and distributed under the conditions of the contract. This means we do not implement a traditional put/call option but rather a tweaked version we describe below. In future work, we will consider counterparty risk broadly and how mitigating it can be combined with our framework to offer more traditional derivatives. 
\item \textbf{Price Feed.} In a derivative where settlement is fully automated, either the underlying security (or a token representing it) needs to be on the blockchain already or the blockchain needs to be able to assign a value to the security---or more precisely, be fed the price it should use in evaluating the code of the contract. In practice, an entity feeding prices (or any external information) into a smart contract is called an oracle. Some related work has examined oracles, and we present our decentralized design in Section~\autoref{price-feed} called PriceGeth, which we have made freely available.\footnote{\PriceGethGit}
\item \textbf{Underlying Financial Model.} The buyer and seller of a derivative, whether implicitly or explicitly, must have some sense of what the probabilistic behaviour of the underlying security must be to determine the terms of the contract. This is the purpose of the infamous Nobel-awarded Black-Scholes model for stock prices --- now obsolete but influential for decades. In our system, such a model is not baked into the functioning of the smart contract but would be used externally to decide favourable terms before buying/selling derivatives. For stocks, modern models (like jump-diffusion) might be used. For derivatives on cryptocurrencies or more esoteric securities, models simply do not exist yet and are an open area of research. Finally, we note that the derivative ultimately settles in Ether and so inflations/deflation of the currency might erode an otherwise profitable derivative.
\end{compactlistn}

In summary, we limit our contributions to (1) and (3) in this work, but also propose this fuller landscape as a useful research agenda for future researchers.

\section{Related Work}

Work on trusted oracles and price feeds, in the Ethereum eco-system, include TownCrier~\cite{zhang2016town} which acts as an attested bridge (running within an SGX enclave) between trusted sources of information and the Ethereum blockchain. Oraclizeit\footnote{\url{http://www.oraclize.it/}} is another price feed which uses the similar workflow to fetch the requested information. Our approach differs from these as PriceGeth publishes the data to the Ethereum blockchain from the trusted source of information and the historical data is available to all smart contracts, however in comparison with the other approaches, is limited to only the published data (Price pairs).

Equibit~\cite{kievit2016equibit} proposes a method to issue, create, disseminate and maintain equity across a broad base of investors without the need of intermediaries for record keeping. It is conceivable that derivative smart-contracts could utilize Equibit equity as payment or settlement method, as opposed to simply using Bitcoin or Ethereum's native digital currencies.

Bentov et al.~\cite{bentov2017decentralized} note than an extension to their work on decentralized prediction markets can be a derivative instruments they call a \textit{capped contracts for difference}. It is similar to the one implemented in Velocity (their paper is not an implementation but a study of game theoretic properties).  

Recent attacks on smart-contracts, such as TheDAO attack~\cite{50millionhack} attracted security researchers to analyze further on this era. Solidity security and survey of the attacks by Atzei et al.~\cite{atzei2016survey} lists some of the known security vulnerabilities and Luu et al. developed a tool for static analysis on smart contract codes~\cite{luu2016making} which we used.

\section{Materials and Methods} \label{materials}

\paragraph*{Smart Contracts.}
A \textit{contract} is a written or spoken agreement between two or more parties that is intended to be enforceable by law. In a \textit{smart contract}, terms are written in code and executed by machines, removing the human performance component (unless if such a component is specified). We can consider our main smart contract as a black box: the inputs are investors' deposited \textit{ether} (Ethereum's cash) and their position on the future price of an asset, either short or long. The smart contract will retain the deposit in escrow and execute a payout calculation and the payout itself when the expiry date comes. The payout is in Ether only, no actual shares are exchanged (a \textit{contract for difference}) and the maximum payout is capped (\textit{limit up/down}). Due to the deposit, there is no counter-party risk however the contract requires a trustworthy price feed and the investors earn zero interest for the duration of the contract. For this reason, we consider this a first step toward more flexible arrangements. The contract disintermediates the trusted role of the exchange (or broker for over-the-counter) and settling/clearing entities. 

\begin{figure}[t!p]
\centering
\includegraphics[width=0.5\textwidth]{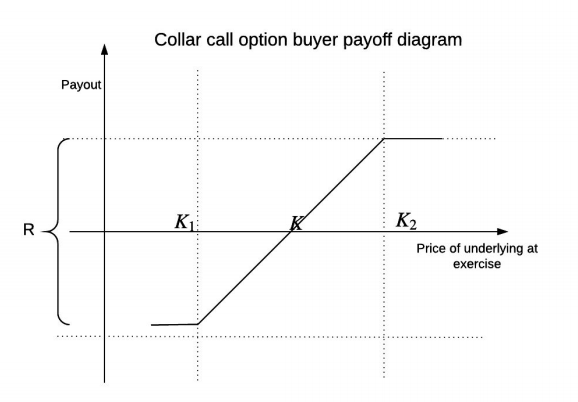}
  \caption{Our collar-esque option with maximum long payout scenario. K1 is the initial price, K2 is the price at expiry time and R is the pre-defined collar for payouts}
\label{collar_options}
\end{figure}

\paragraph{Types of Options.}
We implement a non-standard option that is similar to a collar or hedge wrapper. It is non-standard due to our requirement of escrowing money, which we make to side-step counter-party risk and enable a fully autonomous and disintermediated contract. The contract collects funds from the hedgers/speculators who take opposing positions on the future prospects of an asset: one takes the short position when they believe the underlying asset's value will lose value from its current price, and other takes the opposite long position speculating a rise in the price. In its simplest form, the collar options pay out \$1 for every~\$1 change in the underlying asset (the payout can be made dependent on a drift term or even made non-linear). The payout is limited by the amount of money held in escrow---if the price rises beyond the limit, it is said to be limit up (or limit down in the opposite case) and the payout will be fixed (see Figure~\ref{collar_options}). This kind of payout capping helps the contract holders stay immune to systemic risks and extreme jumps.

\paragraph{Development and Deployment.}
There are a few blockchains that would let us code an autonomous smart contracts: Ethereum, RSK~\cite{RootstockWhitePaper} and more. The decision to work on Ethereum blockchain rather than others solely came from the fact that there are more active developers in the community and maturity of the platform. Even though Ethereum is in early stages, it is more mature than other smart contract compatible platforms. The programming language used for smart contract development is Solidity in most of these platforms. All smart contracts developed and used in this paper has been deployed and tested by our beta testers on Ethereum testnet. In Ethereum blockchain, transactions and processing power costs some small amount of ether called \textit{gas}\footnote{What is gas? \url{http://ethdocs.org/en/latest/contracts-and-transactions/account-types-gas-and-transactions.html\#what-is-gas}}. For each transaction, the sender defines the \textsf{gasLimit} and also \textsf{gasPrice} for processing that transaction and miners decide to include those transactions in the blocks they mine or not. The concept of gas has many angles to discuss which falls outside of the scope of this paper. We will discuss some more in~\autoref{gas-stuff}.



\section{Implementation}

We call our platform \Velocity. We tried to model the real-life scenario of buying an options derivatives. Consider the case where Alice goes to a broker and buys an options contract from Bob. The broker is the one that handles the money transfer and also execute the options contract at the contract expiry time. Now our goal is to replace the broker with a smart contract. For the purpose of a proof of concept, \textbf{the smart contract will also act as Bob}, meaning if Alice buys a short call option, the \Velocity smart contract will put a long call against her short call. This can be generalized so that other entities can fund the contract but for the rest of this paper, \Velocity acts as a market maker. This might lead to users gaming the system, however it's trivial to change the smart contract to wait for the other opponent to enter the contract. We discuss this more in~\autoref{gaming-the-system}. 

\subsection{Velocity Main Smart Contract}
A \Velocity smart contract can be used for speculation on the price of any two assets\footnote{or any other events that an options contract can be based on}, although the Ethereum price is always exposed as the deposits and the withdrawals are done in ETH\footnote{Ethereum symbol}. As for this experiment, we use the price pair of Bitcoin (XBT/BTC) and Ethereum (ETH). If we used price pairs not involving ETH, for example the CAD/USD exchange rate, it would suffice to use two contracts for CAD/ETH and ETH/USD. Or the payout function could be changed to specify how it relates to numbers it is given. Note that in either case, the payout will always be in ETH. In its full generality, any number that changes over time and has a suitable feed (we describe feeds below) can be used: price (stocks, bonds, commodities, etc.), rate (interest, inflation, population, etc.), or something else (average global temperature, number of days without rain, etc). 

\subsubsection{Smart contract.}
The way \Velocity smart contract is implemented, one party purchases a contract by sending a nominal amount of ethereum (0.1 ETH) to the contract's ethereum address. Once confirmed by the network, the contract will fetch a starting price from the price feed, \PriceGeth, and run for a period of time to reach the expiry time. The smart contract would put the same amount of ETH from its pool of funds into escrow for the payout. In the PoC demo, we use 5 ethereum blocks (approximately 1 minute) to settle a contract. When the expiry time reaches, the same party must send another transaction to the contract and call the settlement function to settle the contract which leads to sending the payouts by the smart contract. While this experiment was going under beta testings, we found out that if the user loses the contract, there is no incentive to call the settle function as it would use up some ETH in gas and would not pay the user. This would lead stale money held in the escrow of the smart contract. This made us redesign our settlement functions and write one centralized cron job script to go through the unsettled contracts once a day and call the settle function on the ones that have been expired.

\begin{Verbatim}[commandchars=\\\{\}]
 \PYG{k+kd}{modifier} \PYG{n}{checkMargin}\PYG{p}{(}\PYG{k+kt}{uint} \PYG{n}{amount}\PYG{p}{)} \PYG{p}{\PYGZob{}}
    \PYG{k}{if} \PYG{p}{(}\PYG{n}{amount} \PYG{o}{==} \PYG{p}{(}\PYG{n}{applyLOT}\PYG{p}{(}\PYG{n}{Margin}\PYG{p}{)))}
    \PYG{p}{\PYGZob{}} \PYG{k+kr}{\PYGZus{}} \PYG{p}{;\PYGZcb{}} \PYG{k}{else} \PYG{p}{\PYGZob{}}
        \PYG{n}{Error}\PYG{p}{(}\PYG{l+s+s2}{\PYGZdq{}Invalid Margin!\PYGZdq{}}\PYG{p}{);}
        \PYG{n}{immediateRefund}\PYG{p}{();\PYGZcb{}}
    \PYG{p}{\PYGZcb{}}
\PYG{k+kd}{function} \PYG{n}{goLong}\PYG{p}{()} \PYG{k+kr}{public} \PYG{n}{hasEnoughFunds}\PYG{p}{(}\PYG{n+nb}{msg}\PYG{p}{.}\PYG{n}{value}\PYG{p}{)} \PYG{n}{checkMargin}\PYG{p}{(}\PYG{n+nb}{msg}\PYG{p}{.}\PYG{n}{value}\PYG{p}{)} 
\PYG{n}{payable} \PYG{k}{returns}\PYG{p}{(}\PYG{k+kt}{uint}\PYG{p}{)} \PYG{p}{\PYGZob{}}
    \PYG{n}{lastOptionId} \PYG{o}{=} \PYG{n}{newOption}\PYG{p}{(}\PYG{n+nb}{msg}\PYG{p}{.}\PYG{n}{sender}\PYG{p}{,} \PYG{n+nb}{msg}\PYG{p}{.}\PYG{n}{value}\PYG{p}{,} \PYG{k+kc}{true}\PYG{p}{);}
    \PYG{n}{LongOption}\PYG{p}{(}\PYG{n}{lastOptionId}\PYG{p}{,} \PYG{n+nb}{msg}\PYG{p}{.}\PYG{n}{sender}\PYG{p}{,} \PYG{n+nb}{msg}\PYG{p}{.}\PYG{n}{value}\PYG{p}{,} \PYG{n+nb}{block}\PYG{p}{.}\PYG{n}{number}\PYG{p}{);}
    \PYG{k}{return} \PYG{n}{lastOptionId}\PYG{p}{;}
    \PYG{p}{\PYGZcb{}}
\PYG{n}{Code 1 {:} Velocity Main Smart Contract - Long Option Call, The sender }
       \PYG{n}{of a transaction to goLong() function has to send exactly the}
       \PYG{n}{Margin value and with that he enters the option contract for}
       \PYG{n}{Margin value with the smart Contract}

\end{Verbatim}

\subsubsection{Settle function.} \label{settle-function}
\textsf{exercise()} is responsible in settling the options contract and pay out both parties (see Code 2), in which here is the user and the \Velocity smart contract. Most of the functions are responsible to find the appropriate option contract and calculate the pay outs. However there are some functions that were added later on for security measurements, such as isOpen modifier. Modifiers in Solidity are functions that can check some statements before executing the main function. The first deployed version of \Velocity main contract was vulnerable to a similar (but not the same) attack as the DAO attack, see~\autoref{security}. It was possible for an attacker to call an option contract and upon settling and winning, keep calling the exercise() function using his OptionId and get more of the same amount of payout over and over again. The code was patched and a new smart contract was deployed later in the experiment\footnote{Fix for the multiple payout bug: \url{https://github.com/VelocityMarket/Options-Contract/commit/f3c8d0ef66b886c9ee8b432e92c83f3a4fb525ba}}.
send() is a built-in function in Solidity which handles the sending of funds to other ethereum addresses or contracts. There are known vulnerabilities on how send() function works in solidity which should be appropriately handled. One can use a smart contract address as his option payout address which would execute some code upon receiving any funds and use that code flow to drain the sender's contract. payAndHandle() function tried to use the best security practices to prevent such attacks (see Code 5 for the source code).

\begin{Verbatim}[commandchars=\\\{\}]
  \PYG{k+kd}{modifier} \PYG{n}{isOpen}\PYG{p}{(}\PYG{k+kt}{uint} \PYG{n}{optionId}\PYG{p}{)} \PYG{p}{\PYGZob{}}\PYG{k}{if} \PYG{p}{(}\PYG{n}{AllOptions}\PYG{p}{[}\PYG{n}{optionId}\PYG{p}{].}\PYG{n}{closed}\PYG{p}{)} \PYG{k}{throw}\PYG{p}{;} \PYG{k+kr}{\PYGZus{}} \PYG{p}{;\PYGZcb{}}
  \PYG{k+kd}{function} \PYG{n}{exercise}\PYG{p}{()} \PYG{k+kr}{public} \PYG{p}{\PYGZob{}}
    \PYG{n}{exercise}\PYG{p}{(}\PYG{n}{findOptionId}\PYG{p}{(}\PYG{n+nb}{msg}\PYG{p}{.}\PYG{n}{sender}\PYG{p}{));}
  \PYG{p}{\PYGZcb{}}
  \PYG{k+kd}{function} \PYG{n}{exercise}\PYG{p}{(}\PYG{k+kt}{uint} \PYG{n}{optionId}\PYG{p}{)} \PYG{k+kr}{public} \PYG{n}{isOpen}\PYG{p}{(}\PYG{n}{optionId}\PYG{p}{)} \PYG{k}{returns}\PYG{p}{(}\PYG{k+kt}{bool}\PYG{p}{)} \PYG{p}{\PYGZob{}}
    \PYG{c+c1}{// REMOVED SOME CODE TO SAVE SPACE, FULL SOURCE CODE ON  GITHUB}
    \PYG{n}{AllOptions}\PYG{p}{[}\PYG{n}{optionId}\PYG{p}{].}\PYG{n}{closed} \PYG{o}{=} \PYG{k+kc}{true}\PYG{p}{;} \PYG{c+c1}{//before payouts to prevent replay attacks}
    \PYG{n}{LockedBalance} \PYG{o}{\PYGZhy{}=} \PYG{n}{AllOptions}\PYG{p}{[}\PYG{n}{optionId}\PYG{p}{].}\PYG{n}{amount}\PYG{p}{;} \PYG{c+c1}{//release escrow}
    \PYG{c+c1}{// Payout calculation}
    \PYG{k}{if} \PYG{p}{(}\PYG{n}{pricesToCheck}\PYG{p}{.}\PYG{n}{pricediff} \PYG{o}{\PYGZgt{}=} \PYG{p}{(}\PYG{k+kt}{int}\PYG{p}{(}\PYG{n}{Margin}\PYG{p}{)))} \PYG{p}{\PYGZob{}} \PYG{c+c1}{// Pay Long}
        \PYG{c+c1}{//pay long}
        \PYG{k}{return} \PYG{n}{payAndHandle}\PYG{p}{(}\PYG{n}{optionId}\PYG{p}{,} \PYG{n}{AllOptions}\PYG{p}{[}\PYG{n}{optionId}\PYG{p}{].}\PYG{n}{Long}\PYG{p}{,} 
\PYG{l+m+mi}{2} \PYG{o}{*} \PYG{n}{AllOptions}\PYG{p}{[}\PYG{n}{optionId}\PYG{p}{].}\PYG{n}{amount}\PYG{p}{);}
    \PYG{p}{\PYGZcb{}}
    \PYG{k}{if} \PYG{p}{((}\PYG{l+m+mi}{0} \PYG{o}{\PYGZlt{}} \PYG{n}{pricesToCheck}\PYG{p}{.}\PYG{n}{pricediff}\PYG{p}{)} \PYG{o}{\PYGZam{}\PYGZam{}} \PYG{p}{(}\PYG{n}{pricesToCheck}\PYG{p}{.}\PYG{n}{pricediff} \PYG{o}{\PYGZlt{}} \PYG{p}{(}\PYG{k+kt}{int}\PYG{p}{(}\PYG{n}{Margin}\PYG{p}{))))} \PYG{p}{\PYGZob{}}
      \PYG{k}{return} \PYG{p}{(}\PYG{n}{payAndHandle}\PYG{p}{(}\PYG{n}{optionId}\PYG{p}{,} \PYG{n}{AllOptions}\PYG{p}{[}\PYG{n}{optionId}\PYG{p}{].}\PYG{n}{Long}\PYG{p}{,} 
        \PYG{p}{(}\PYG{n}{AllOptions}\PYG{p}{[}\PYG{n}{optionId}\PYG{p}{].}\PYG{n}{amount} \PYG{o}{+} \PYG{n}{pricesToCheck}\PYG{p}{.}\PYG{n}{priceDiffLOT}\PYG{p}{))} \PYG{o}{\PYGZam{}\PYGZam{}} 
               \PYG{n}{payAndHandle}\PYG{p}{(}\PYG{n}{optionId}\PYG{p}{,} \PYG{n}{AllOptions}\PYG{p}{[}\PYG{n}{optionId}\PYG{p}{].}\PYG{n}{Short}\PYG{p}{,} 
               \PYG{p}{(}\PYG{n}{AllOptions}\PYG{p}{[}\PYG{n}{optionId}\PYG{p}{].}\PYG{n}{amount} \PYG{o}{\PYGZhy{}} \PYG{n}{pricesToCheck}\PYG{p}{.}\PYG{n}{priceDiffLOT}\PYG{p}{)));}
    \PYG{p}{\PYGZcb{}}
  \PYG{p}{\PYGZcb{}}
\PYG{n}{Code 2{: }Settle function of main options contract}
\end{Verbatim}

\subsubsection{Source Code}
API documentation for other smart contracts to use the functionality and also Python and NodeJS clients to communicate with the main smart contract are available on Github\footnote{Simple collared option smart contract: \url{https://github.com/VelocityMarket/Options-Contract}}.



\subsection{Price feed} 
\label{price-feed}
\begin{figure}[t!p]
\centering
\includegraphics[width=0.8\linewidth]{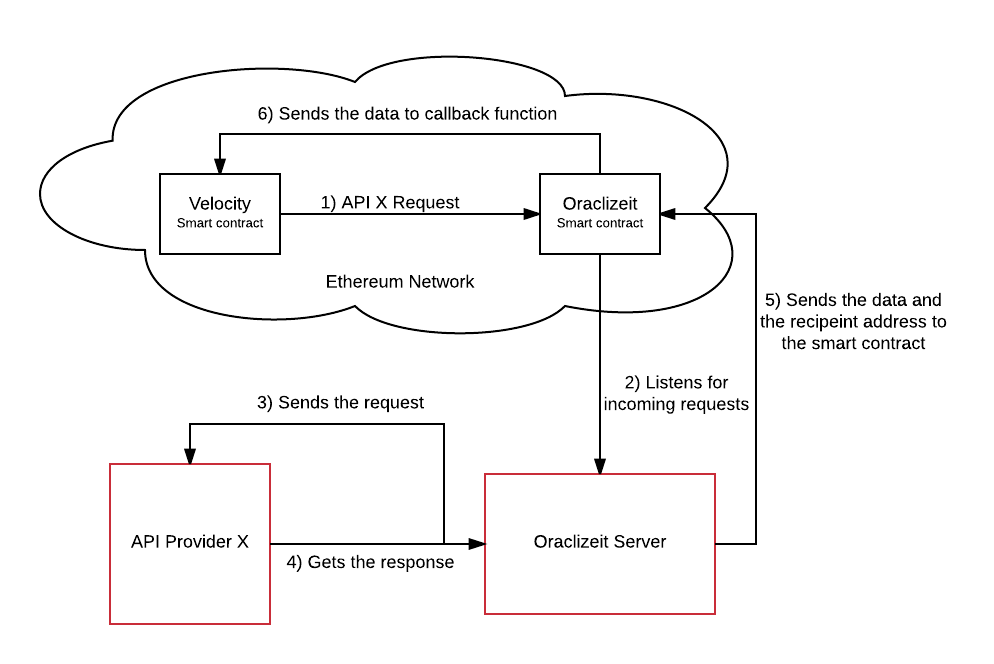}
  \caption{Oraclizeit work flow}
\label{oraclizeit-flow}
\end{figure}

A decentralized Price feed is an essential requirement for having a decentralized derivative market. There are a few proposals on how to fetch the price in a smart contract. One is using \textit{Smart Contract} oracles\footnote{Data and Payments for your Smart Contracts \url{https://smartcontract.com/}}, they offer daily updates for the price using a predefined data source. This was not an option to be used for our purpose as a daily update is not sufficient for short term derivative markets. Another option that could be used was \Oraclizeit. They way \Oraclizeit works is that the client smart contract, \Velocity main contract in our case, sends a transaction to \Oraclizeit smart contract with the required API url and the fields it needs, sometime after the confirmation by the network, \Oraclizeit smart contract sends a callback transaction to \Velocity smart contract with the requested data (\autoref{oraclizeit-flow}).

For the first implementation of \Velocity smart contract we used \Oraclizeit method to fetch the price.

As mentioned before, most of the decentralized application infrastructure on Ethereum blockchain are in Beta state and might not work as intended. This applies for \Oraclizeit, specially as by design they have a central server which can stop working without any notice or visible signs. The red boxes in~\autoref{oraclizeit-flow} indicates the centralized parts of the system. As you can see in Code 3, \Oraclizeit will send the price to the callback function at the time of the call and also execute the exercise() function which is responsible for saving the price and calculating the payout amounts. This makes the callback function one of the important functions which should be called at the specific time. 
\begin{Verbatim}[commandchars=\\\{\}]
\PYG{n}{oraclize\PYGZus{}setProof}\PYG{p}{(}\PYG{n}{proofType\PYGZus{}TLSNotary} \PYG{o}{|} \PYG{n}{proofStorage\PYGZus{}IPFS}\PYG{p}{);}
\PYG{c+c1}{//oraclize\PYGZus{}setNetwork(2); //}
\PYG{n}{priceUrl} \PYG{o}{=} \PYG{l+s+s2}{\PYGZdq{}json(https://www.bitstamp.net/api/v2/ticker/btcusd).last\PYGZdq{}}\PYG{p}{;}
\PYG{k+kd}{function} \PYG{n}{updateBTCUSDFromFeed}\PYG{p}{(}\PYG{k+kt}{uint} \PYG{n}{delay}\PYG{p}{)\PYGZob{}}
  \PYG{n}{oraclize\PYGZus{}query}\PYG{p}{(}\PYG{n}{delay}\PYG{p}{,} \PYG{l+s+s2}{\PYGZdq{}URL\PYGZdq{}}\PYG{p}{,}
    \PYG{n}{priceUrl}\PYG{p}{,} \PYG{l+m+mi}{400000}\PYG{p}{);} 
  \PYG{p}{\PYGZcb{}}
\PYG{k+kd}{function} \PYG{n}{\PYGZus{}\PYGZus{}callback}\PYG{p}{(}\PYG{k+kt}{bytes32} \PYG{n}{myid}\PYG{p}{,} \PYG{k+kt}{string} \PYG{n}{result}\PYG{p}{,} \PYG{k+kt}{bytes} \PYG{n}{proof}\PYG{p}{)} \PYG{p}{\PYGZob{}}
  \PYG{k}{if} \PYG{p}{(}\PYG{n+nb}{msg}\PYG{p}{.}\PYG{n}{sender} \PYG{o}{!=} \PYG{n}{oraclize\PYGZus{}cbAddress}\PYG{p}{())} \PYG{k}{throw}\PYG{p}{;}
    \PYG{k+kt}{uint} \PYG{n}{BTCUSDFeed}\PYG{p}{;}
    \PYG{n}{BTCUSDFeed} \PYG{o}{=} \PYG{n}{parseInt}\PYG{p}{(}\PYG{n}{result}\PYG{p}{,} \PYG{l+m+mi}{2}\PYG{p}{);}
  \PYG{n}{exercise}\PYG{p}{()} \PYG{c+c1}{// this function exercises the contract to calculate the payouts}
  \PYG{p}{\PYGZcb{}}
\PYG{n}{Code 3 {:} Implementation of Oraclizeit price feed in Velocity smart contract}

\end{Verbatim}

In our testing period, we encountered multiple problems with this design:

\begin{enumerate}

\item The callback would not happen at all, which would result in an unsettled options contract. \Oraclizeit support team were helpful and fixed this issue later on.

\item The callback would happen with some delays, which would result in inconsistency in the fetched price with the the options contract expiry date. decentralized networks have some latency by design, realtime does not really mean anything in such networks, hence counting on a transaction to happen at a exact time is not the best solution.

\item The callback would happen with insufficient gas, which would result in the failure to properly run exercise() function and thus failiure to settle the options contract. \Oraclizeit library offers a way to send more gas than needed in case the callback function needs more gas, however on the time of this experiment that functionality was not working properly.
\end{enumerate}

\begin{figure}[t!p]
\centering
\includegraphics[width=0.8\linewidth]{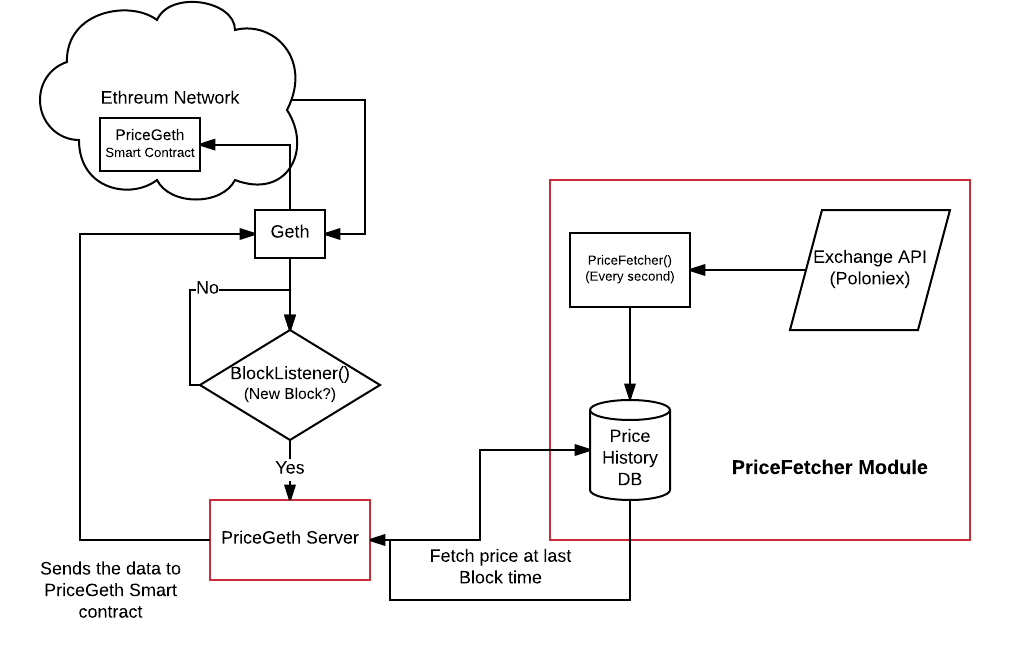}
  \caption{PriceGeth Work Flow}
\label{pricegeth-flow}
\end{figure}

\paragraph{PriceGeth} We designed \PriceGeth\footnote{Price API for Smart-Contracts on Ethereum Blockchain  \PriceGethGit} to publish (almost) realtime price pairs to Ethereum blockchain. 
This is how \PriceGeth works (also see~\autoref{pricegeth-flow}):
\begin{enumerate}
\item PriceFetcher server is saving an exchange Prices (USDBTC, BTCETH, BTCETC, BTCDOGE) every 1 second in a database
\item BlockListener is listening on using Geth\footnote{Official Go implementation of the Ethereum protocol \url{ https://geth.ethereum.org }} for new blocks
\item When BlockListener sees a new block it fetches the price at the Blocktime from PriceFetcher Module
\item \PriceGeth server sends the data to \PriceGeth smart contract (Code 4) and updates the latest price
\end{enumerate}

\PriceGeth smart contract would keep all the historical prices and all would be available to all smart contracts on Ethereum blockchain for free (no gas needed to fetch the price). The reason this is almost realtime, goes back to the nature of blockchains. Time units as in seconds and minutes are not meaningful for most of the blockchain applications, but the block height can be used as the time unit, meaning the time of each block is known to all users of the blockchain, but before a block is published no other time units can be used. This is why we designed PriceFetcher module to connect to an exchange API and saves the price pairs every second, to have the price for the previous block time anytime a new Ethereum block is generated. 

%
\begin{Verbatim}[commandchars=\\\{\}]
\PYG{k+kd}{struct} \PYG{n}{Feed} \PYG{p}{\PYGZob{}}
      \PYG{k+kt}{uint}    \PYG{n}{USDBTC}\PYG{p}{;}
      \PYG{k+kt}{uint40}    \PYG{n}{BTCETH}\PYG{p}{;}
      \PYG{k+kt}{uint40}    \PYG{n}{BTCETC}\PYG{p}{;}
      \PYG{k+kt}{uint40}    \PYG{n}{BTCDOGE}\PYG{p}{;}
      \PYG{k+kt}{uint40}  \PYG{n}{timestamp}\PYG{p}{;}
      \PYG{k+kt}{uint}    \PYG{n}{blockNumber}\PYG{p}{;}
    \PYG{p}{\PYGZcb{}}
\PYG{k+kd}{mapping} \PYG{p}{(}\PYG{k+kt}{uint} \PYG{o}{=\PYGZgt{}} \PYG{n}{Feed}\PYG{p}{)} \PYG{n}{priceHistory}\PYG{p}{;}
\PYG{k+kd}{function} \PYG{n}{setPrice}\PYG{p}{(}\PYG{k+kt}{uint40} \PYG{n}{timestamp}\PYG{p}{,} \PYG{k+kt}{uint40} \PYG{n}{blocknumber}\PYG{p}{,} \PYG{k+kt}{uint} \PYG{n}{USDBTC}\PYG{p}{,} 
\PYG{k+kt}{uint40} \PYG{n}{BTCETH}\PYG{p}{,} \PYG{k+kt}{uint40} \PYG{n}{BTCETC}\PYG{p}{,} \PYG{k+kt}{uint40} \PYG{n}{BTCDOGE}\PYG{p}{)} \PYG{n}{ifOwner}\PYG{p}{()} \PYG{p}{\PYGZob{}}
  \PYG{k}{if} \PYG{p}{(}\PYG{n}{firstBlock} \PYG{o}{==} \PYG{l+m+mi}{0}\PYG{p}{)} \PYG{n}{firstBlock} \PYG{o}{=} \PYG{n}{blocknumber}\PYG{p}{;}
  \PYG{n}{priceHistory}\PYG{p}{[}\PYG{n}{lastBlock}\PYG{p}{].}\PYG{n}{timestamp} \PYG{o}{=} \PYG{n}{timestamp}\PYG{p}{;} 
  \PYG{n}{priceHistory}\PYG{p}{[}\PYG{n}{lastBlock}\PYG{p}{].}\PYG{n}{blockNumber} \PYG{o}{=} \PYG{n}{blocknumber}\PYG{p}{;} 
  \PYG{n}{priceHistory}\PYG{p}{[}\PYG{n}{lastBlock}\PYG{p}{].}\PYG{n}{USDBTC} \PYG{o}{=} \PYG{n}{USDBTC}\PYG{p}{;}
  \PYG{n}{priceHistory}\PYG{p}{[}\PYG{n}{lastBlock}\PYG{p}{].}\PYG{n}{BTCETH} \PYG{o}{=} \PYG{n}{BTCETH}\PYG{p}{;}
  \PYG{n}{priceHistory}\PYG{p}{[}\PYG{n}{lastBlock}\PYG{p}{].}\PYG{n}{BTCETC} \PYG{o}{=} \PYG{n}{BTCETC}\PYG{p}{;}
  \PYG{n}{priceHistory}\PYG{p}{[}\PYG{n}{lastBlock}\PYG{p}{].}\PYG{n}{BTCDOGE} \PYG{o}{=} \PYG{n}{BTCDOGE}\PYG{p}{;}
  \PYG{n}{PriceUpdated}\PYG{p}{(}\PYG{n}{timestamp}\PYG{p}{,} \PYG{n}{blocknumber}\PYG{p}{,} \PYG{n}{USDBTC}\PYG{p}{,} \PYG{n}{BTCETH}\PYG{p}{,} \PYG{n}{BTCETC}\PYG{p}{,} \PYG{n}{BTCDOGE}\PYG{p}{);}
\PYG{p}{\PYGZcb{}}
\PYG{n}{Code 4 {:} Pricegeth Main Smart contract}

\end{Verbatim}

\PriceGeth is a proof of concept implementation of having a trusted entity publishing price pairs to the blockchain and we are aware of the implications of trusting the \textsf{PriceFetcher} not to manipulate the prices. \textsf{PriceFetcher} is the central point of failure in \PriceGeth design and should be addressed in future work. However after further research, it is almost impossible to have a truly trustless decentralized price feed unless we have a decentralized exchange infrastructure on the blockchain. This exchange can be used as the price oracle as the order books would be stored on the blockchain and hence there is no one single point of trust. The red boxes in ~\autoref{pricegeth-flow} are indicating the centralized parts of this implementation.
\PriceGeth is released as a stand alone smart contract and also a library to be used in other smart contracts to use the price feed free of charge\footnote{PriceGeth Library \PriceGethGit}. 
Another challenge of \PriceGeth design is that \textsf{PricePublisher} is paying the gas for publishing and storing all the price pairs, and as there is no incentive of doing so, it is not an inefficient way of offering price oracles. \PriceGeth can be implemented in a way that clients should use a token issued to them beforehand to fetch the price, or require payments to release the price data. 

By design \PriceGeth operator should not be able to use \Velocity options as he can manipulate the price to game the system.

There is a similar work on price feeds titled Town Crier~\cite{zhang2016town}, which uses TLS security to prove the fact that the data sent to the smart contract is exactly as the one provided by the API, conceptually similar to \Oraclizeit TLSNotary-proof\footnote{\url{https://docs.oraclize.it/\#security-tlsnotary-proof}}. TownCrier uses Intel SGX in their central server which insures the integrity of hardware used and thus insures no manipulation is done on the server. Even though one can argue that the data provider is a trusted entity, one of the goals to have a decentralized application is to have no trusted entity in the infrastructure and to have a trustless system.


\section{Discussion} \label{discussion}

\subsubsection{Security} \label{security}
Smart contracts have introduced some new security concerns to developers. Notions like gas usage and consensus and most importantly a function that pays out irreversible money are new to most of the developers hence the ability to develop a secure smart contract is hard to grasp. One of the visible examples of security issues is the attack on The DAO,  Decentralized Autonomous Organization\footnote{\url{https://github.com/slockit/DAO}}. The goal of the DAO was to remove all the need for any venture capital intervention or any other third party for fundraising on a new idea or a company through crowdfunding and giving the investors tokens (shares) of the company. However due to an issue splitDAO function which was responsible to manage and fund new child DAOs or projects, an attacker was able to take one third of the money in the original DAO, worth approximately 86 million USD~\cite{DAOexploit} at the time of the attack, this vulnerability is dubbed \textit{Reentrancy Vulnerability}. 

\begin{figure}[t!p]
\centering
\includegraphics[width=0.4\linewidth]{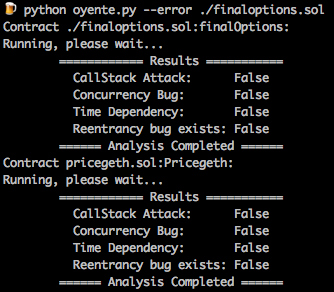}
  \caption{Results of Smart Contract analysis tool called Oyente~\cite{luu2016making} to find security bugs}
\label{oyente_result}
\end{figure}

Luu et al.~\cite{luu2016making} developed a symbolic execution tool called ``Oyente'' to find potential security bugs, which they proved effective by running on Ethereum blockchain and successfully identifying The DAO vulnerability. We used this tool to analyze our code (see~\autoref{oyente_result}).

Another family of vulnerabilities that have caused some of the known attacks are \textit{Mishandled Exceptions}, which mostly has caused Denial of Service attacks on individual smart contracts. In \Velocity main contract we used \textit{modifier} functions to sanitize the inputs to narrow down the probability of such exceptions. Another set of attacks \textit{Timestamp Dependence} and \textit{Transaction-Ordering Dependence} are interesting to ponder, however due to the design of \Velocity and \PriceGeth, they are not applicable to these smart contracts. As an example, usage of timestamp was replaced by Ethereum blocknumber and smart contracts time is based on the block number rather than seconds and minutes. 
There has been more security bugs in solidity compiler, a few related bugs were explained in~\ref{settle-function}.

\begin{Verbatim}[commandchars=\\\{\}]
    \PYG{k+kd}{function} \PYG{n}{payAndHandle}\PYG{p}{(}\PYG{k+kt}{uint} \PYG{n}{optionId}\PYG{p}{,} \PYG{k+kt}{address} \PYG{n}{addr}\PYG{p}{,} \PYG{k+kt}{uint} \PYG{n}{amount}\PYG{p}{)}
 \PYG{k+kr}{private} \PYG{k}{returns} \PYG{p}{(}\PYG{k+kt}{bool} \PYG{n}{success}\PYG{p}{)} \PYG{p}{\PYGZob{}}
      \PYG{k}{if} \PYG{p}{(}\PYG{n}{addr}\PYG{p}{.}\PYG{n+nb}{send}\PYG{p}{(}\PYG{n}{amount}\PYG{p}{))} \PYG{p}{\PYGZob{}}
            \PYG{n}{optionPaid}\PYG{p}{(}\PYG{n}{optionId}\PYG{p}{,} \PYG{n}{addr}\PYG{p}{,} \PYG{n}{amount}\PYG{p}{);} \PYG{c+c1}{//event for successful payment}
      \PYG{p}{\PYGZcb{}} \PYG{k}{else} \PYG{p}{\PYGZob{}} \PYG{k}{throw}\PYG{p}{;\PYGZcb{}}
      \PYG{k}{return} \PYG{k+kc}{true}\PYG{p}{;}
    \PYG{p}{\PYGZcb{}}
\PYG{n}{Code 5 {:} Secure payouts in smart contracts}
\end{Verbatim}

\subsubsection{Gas Sustainability}\label{gas-stuff}
The concept of gas usage for processing power is not easy to grasp even for long term developers. People might be familiar with limited computational or storage resources, but the concept of passing gasLimit to a function to use to process inputs is a new concept. Each step has its own estimated gas usage, as an example to store a value in a variable, you have to pay \textit{100 Wei}\footnote{Wei: Smallest unit of Ethereum, equevalent to 0.000000000000000001 ETH} for each \textit{sstore} call\footnote{put into permanent storage}. This should be considered that there's a cap for gas usage for each transaction and block, thus complex computation should be split into multiple transactions which makes smart contract design more complicated than they are. Also we should mention that function calls can fail due to the fact that they run out of gas and they don't have enough gas to finish their required computation or storage. This can cause unpredicted behaviour from the smart contract as there would be broken flows in the code which should have been handled by the developer. The gas usage could change as there are updates and security patches to Ethereum protocol, e.g transaction spam attack\footnote{Long-term gas cost changes for IO-heavy operations to mitigate transaction spam attacks \url{https://github.com/ethereum/EIPs/issues/150}}. It might take multiple implementation of the same function to find an equilibrium between readability and gas efficiency.  

\subsubsection{Misuse of the contract} \label{gaming-the-system}
In the current implementation of \Velocity smart contract, one can call the Long option when he is sure of the price increase between the start time and expiry time and keep on doing this until there is no money left in the smart contract's pool of funds. This is because the smart contract calls the opposite of the incoming option call blindly. However in future work, there should be market scoring rule which depends on how many short option calls are placed comparing to the long calls and make it more expensive to call short when there are more short option calls than long calls. 

\subsubsection{Collar Option library}
\Velocity smart contract can be used as a module in any other smart contract to handle option calls and execute some functions on the expiry time. This smart contract was written as a proof of concept and was released under \textit{GPL} license\footnote{\url{https://github.com/VelocityMarket/Options-Contract}}. 


\section{Future work}
As discussed in~\autoref{price-feed}, fully decentralized Price feeds and oracles are needed in order to have a trustless decentralized financial market. This can be done by having a decentralized exchange to extract prices from using smart contracts. Even though there has been many price feed methods discussed, none of them seem to have trustless infrastructure. 
Smart contracts security is not well practiced and there are many unknown attack vectors in the eco system, from solidity compiler security bugs~\cite{Reitwiessner2016ethereum} to best practice security implementations~\cite{ContractSecurityTechniques}, there is work to be done and tests to have a more mature secure eco-system to work with, Specially if the end goal is to have a decentralized financial application in place where money is at stake. 

As for the options contracts, there should be more research and work on the payouts to make them smarter. One proposed solution is to have market scoring rules in place, which means if there are more open short option calls than long calls, it should get more expensive to call short options and vice-versa. Smart contracts are unchangeable piece of code that run autonomously, meaning if there's a market crash or systematic error, there cannot be anything to do to suspend the payouts and shut down the application, unless with pre-defined functions in the smart contract which only the owner can trigger, which would be a double standard in the trustless eco-system. 


\section{Conclusion}

Even though the idea of having a fully autonomous and decentralized derivative market is intriguing, the infrastructure to reach this goal is still missing from the underlying network. As for example, price feed is one of the essentials of such a market and it should be done in a fully decentralized trustless way to prevent fraud and market manipulation by the feed provider. All the existing solutions today, have a central point that can manipulate data, it is either the exchange API or the component responsible to publish the price. As discussed before, one of the only solutions to this problem is to have a fully decentralized exchange on the network to provide realtime price feed for other smart contracts. There are some work done on decentralized exchanges~\cite{clark2014decentralizing}, although there is no real world deployment of such a system at the time of writing. 
Smart contracts are fascinating idea that can revolutionize the technology by removing the middlemen, however the underlying technology is more on the proof of concept level than mature enough to be used on the real world scenarios. We should also mention that the barrier for people to have the relevant crypto-currency to work with such systems still exists.

\bibliography{bib/bib}

\newpage
\appendix

\section{Demo Website (UI) for the Velocity smart contract}

\begin{figure}[htb!p]
\centering
\includegraphics[width=\linewidth]{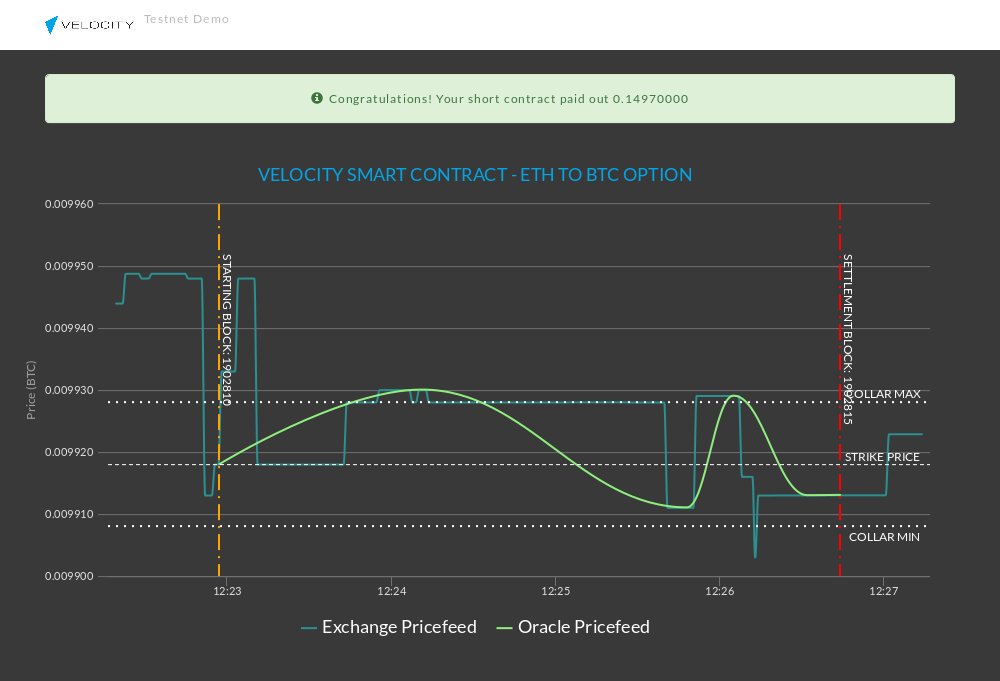}
  \caption{Velocity Options Smart Contract Demo}
\label{velocity_us_demo}
\end{figure}

\end{document}